\documentclass[%
 reprint,
 amsmath,amssymb,
 aps,
 prl,
]{revtex4-1}

\usepackage{graphicx}
\usepackage[space]{grffile}
\usepackage{latexsym}
\usepackage{textcomp}
\usepackage{longtable}
\usepackage{multirow,booktabs}
\usepackage{amsfonts,amsmath,amssymb}
\usepackage{natbib}
\usepackage{url}
\usepackage{hyperref}
\hypersetup{colorlinks=false,pdfborder={0 0 0}}
\newif\iflatexml\latexmlfalse
\DeclareGraphicsExtensions{.pdf,.PDF,.png,.PNG,.jpg,.JPG,.jpeg,.JPEG}

\usepackage[utf8]{inputenc}
\usepackage[english]{babel}
\usepackage{color}

\begin{document}

\title{Time-resolved multi-mass ion imaging: femtosecond UV-VUV pump-probe spectroscopy with the PImMS camera}

\author{Ruaridh Forbes}
\affiliation{Department of Physics and Astronomy, University College London, Gower Street, London, WC1E 6BT, United Kingdom}
\affiliation{Department of Physics, University of Ottawa, 150 Louis Pasteur, Ottawa, ON, K1N 6N5, Canada}

\author{Varun Makhija}
\affiliation{Department of Physics, University of Ottawa, 150 Louis Pasteur, Ottawa, ON, K1N 6N5, Canada}

\author{K\'evin Veyrinas}
\affiliation{Department of Physics, University of Ottawa, 150 Louis Pasteur, Ottawa, ON, K1N 6N5, Canada}

\author{Albert Stolow}
\affiliation{Department of Physics, University of Ottawa, 150 Louis Pasteur, Ottawa, ON, K1N 6N5, Canada}
\affiliation{Department of Chemistry, University of Ottawa, 10 Marie Curie, Ottawa, Ontario, K1N 6N5, Canada}
\affiliation{National Research Council of Canada, 100 Sussex Drive, Ottawa, Ontario K1A 0R6, Canada}

\author{Jason W. L. Lee}
\affiliation{Chemistry Research Laboratory, Department of Chemistry, University of Oxford, 12 Mansfield Road, Oxford OX1 3TA, United Kingdom}

\author{Michael Burt}
\affiliation{Chemistry Research Laboratory, Department of Chemistry, University of Oxford, 12 Mansfield Road, Oxford OX1 3TA, United Kingdom}

\author{Mark Brouard}
\affiliation{Chemistry Research Laboratory, Department of Chemistry, University of Oxford, 12 Mansfield Road, Oxford OX1 3TA, United Kingdom}

\author{Claire Vallance}
\affiliation{Chemistry Research Laboratory, Department of Chemistry, University of Oxford, 12 Mansfield Road, Oxford OX1 3TA, United Kingdom}

\author{Iain Wilkinson}
\affiliation{National Research Council of Canada, 100 Sussex Drive, Ottawa, Ontario K1A 0R6, Canada}
\affiliation{Methods for Material Development, Helmholtz-Zentrum Berlin, Hahn-Meitner-Platz 1, 14109 Berlin, Germany}

\author{Rune Lausten}
\affiliation{National Research Council of Canada, 100 Sussex Drive, Ottawa, Ontario K1A 0R6, Canada}

\author{Paul Hockett}
\affiliation{National Research Council of Canada, 100 Sussex Drive, Ottawa, Ontario K1A 0R6, Canada}


\begin{abstract}
The Pixel-Imaging Mass Spectrometry (PImMS) camera allows for 3D charged particle imaging measurements, in which the particle time-of-flight is recorded along with $(x,y)$ position. Coupling the PImMS camera to an ultrafast pump-probe velocity-map imaging spectroscopy apparatus therefore provides a route to time-resolved multi-mass ion imaging, with both high count rates and large dynamic range, thus allowing for rapid measurements of complex photofragmentation dynamics. Furthermore, the use of vacuum ultraviolet wavelengths for the probe pulse allows for an enhanced observation window for the study of excited state molecular dynamics in small polyatomic molecules having relatively high ionization potentials. Herein, preliminary time-resolved multi-mass imaging results from C$_2$F$_3$I photolysis are presented. The experiments utilized femtosecond UV and VUV (160.8~nm and 267~nm) pump and probe laser pulses in order to demonstrate and explore this new time-resolved experimental ion imaging configuration. The data indicates the depth and power of this measurement modality, with a range of photofragments readily observed, and many indications of complex underlying wavepacket dynamics on the excited state(s) prepared.
\end{abstract}

\maketitle

\section{Introduction}
The recently developed Pixel-Imaging Mass Spectrometry (PImMS) camera provides a hardware platform with pixel-addressable imaging \cite{Nomerotski_2010,Clark_2012,Wilman_2012}, an enabling technology for time-resolved or ``3D" ion imaging. In this context, each event recorded by the camera is a function of $(x,y,t)$, where $(x,y)$ defines the spatial coordinates and $t$ is the time of the event within the exposure. The PImMS camera was primarily developed for photofragment imaging mass spectroscopy in combination with a velocity mapping ion lens and a conventional MicroChannel Plate (MCP)/phosphor-based imaging detector. In a time-of-flight experiment, the ability to record arrival times of particles at the detector enables imaging of multiple mass fragments during each experimental cycle, with the time-resolution defined by the clock-speed of the camera. The camera offers fast data acquisition times and high accuracy for complex fragmentation patterns, and - under optimal conditions - the time resolution can also be exploited for slice imaging modalities. This is in contrast to conventional ``2D" imaging, in which only $(x,y)$ data is obtained by a Charge-Coupled Device (CCD)-camera-based detection system. In such systems, additional time-gating methods are required for fast timing applications, such as pulsing the MCP or an image intensifier coupled to the CCD chip. This typically restricts data acquisition to a small temporal window of interest, e.g. a single mass fragment, necessitating significant experimental effort for multi-mass imaging studies since each fragment must be imaged independently. While other methods of 3D imaging are in use, with delay-line based methods a particularly widespread example \cite{Moshammer_1996,Ullrich_1997,Davies_1999,Lafosse_2000, Gebhardt_2001, Strasser_2000}, the PImMS sensor, as well as other fast frame cameras \cite{Lee_2014}, offers the ability to run at significantly higher count-rates, so is ideal for multi-mass imaging and/or experiments with low repetition rates and/or a large dynamic range of mass channel yields. Particularly with the advent of x-ray free electron lasers with low repetition rates, as well as tabletop extreme ultra-violet/soft x-ray sources that offer relatively low photon fluences, these experimental capabilities provide a unique tool for time-resolved ion imaging experiments. 

To date, the PImMS camera has been demonstrated in a range of applications, including multi-mass imaging \cite{Clark_2012}, 3D ion imaging \cite{Amini_2015}, Coulomb-explosion imaging of aligned molecules \cite{Slater_2014,Slater_2015} and for time-resolved ion imaging experiments at the FLASH free-electron laser \cite{Rolles2016}. Herein, preliminary work combining the PImMS camera with a velocity-map imaging (VMI) spectrometer coupled to a femtosecond (fs) vacuum ultra-violet (VUV) source is described, demonstrating fs time-resolved multi-mass ion imaging. The VUV source is based on the non-collinear four-wave mixing design of Noack and coworkers \cite{Beutler_2010,Ghotbi_2010} and, in our configuration, provides pulses centered at 160.8~nm with a bandwidth of $\approx$~1.5~nm and duration of $\approx$~40~fs. For pump-probe molecular dynamics experiments, fs VUV sources offer hard photons for pump and/or probe, hence an extended observation window for vibronic wavepacket dynamics as compared to lower photon energies. The benefit of VUV in this regard has been demonstrated by Radloff and co-workers for time-resolved mass spectrometry studies of internal conversion and photodissociation of small polyatomic molecules \cite{Farmanara_1999}, and more recently has been applied to time-resolved photoelectron imaging by Suzuki and co-workers \cite{Fuji_2011,Kobayashi_2015}. Here we apply a similar methodology to time-resolved photofragmentation studies. 

The initial experiments discussed herein utilized a standard two-pulse pump-probe spectroscopy configuration, in which the VUV pulses were combined with 266~nm pulses in the interaction region of a Velocity Map Imaging (VMI) spectrometer \cite{Eppink_1997,Parker_1997}, and pump-probe time-resolved ion velocity-map images were obtained by the PImMS camera as a function of pump-probe delay $\Delta t$. The experiments were aimed at exploring the capabilities of the PImMS camera for fs time-resolved ion imaging in the experimental configuration described, and additionally permitted us to carry out a survey of the fs dynamics for a range of small polyatomic species following excitation with VUV light. With limited experimental time available, these studies were planned to provide a foundation for further, longer experimental timescale, precision measurements; with this aim in mind, the experimental methodology and preliminary results are reported herein, but more detailed analysis is deferred for follow-up work.

\section{Experimental Details}

\subsection{Optical setup \& VUV source}
\begin{figure*}
\begin{center}
\includegraphics[width=1\textwidth]{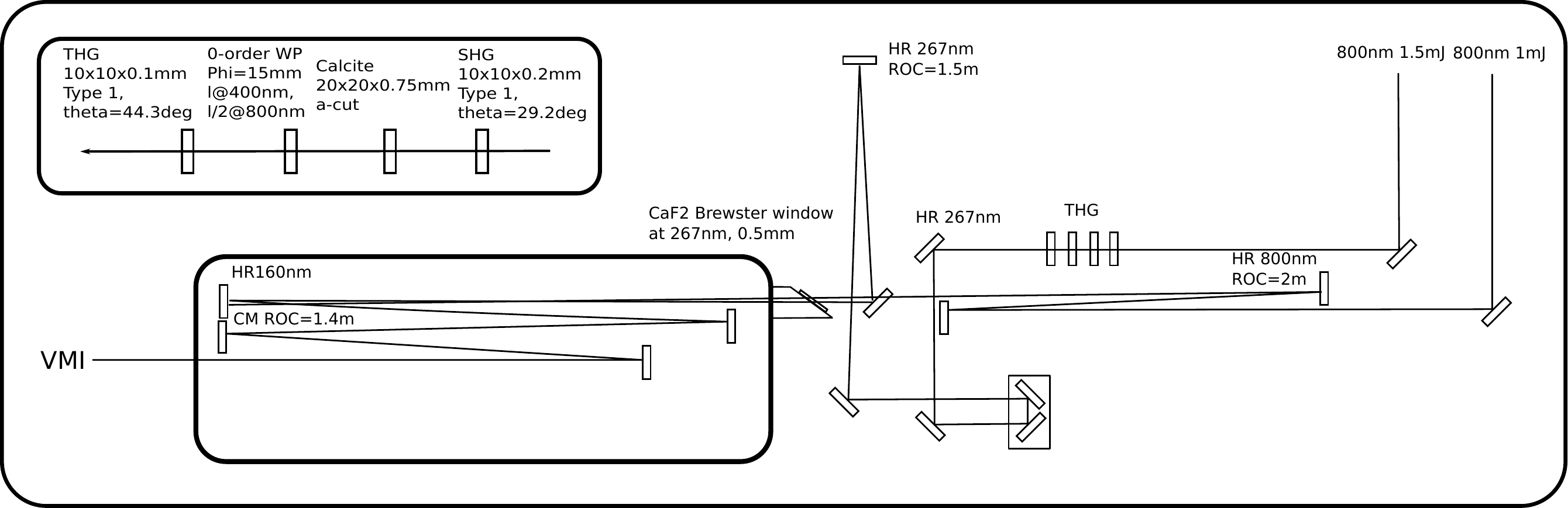}
\caption{{\label{fig:VUV_generation} Schematic overview of the optical set-up used in generation of the 160.8~nm (pump) pulses. The 267~nm (130~$\mu$J) and 800~nm (1~mJ) pulses were focused non-collinearly into a vacuum chamber held at a pressure of 32~Torr in Ar. The generating colors ($\omega$ and 3$\omega$) were separated from the VUV light using dichroic mirrors with high reflectivity at 160~nm and high transmission at 267~nm and 800~nm.%
}}
\end{center}
\end{figure*}

\begin{figure}
\begin{center}
\includegraphics[width=1.0\columnwidth]{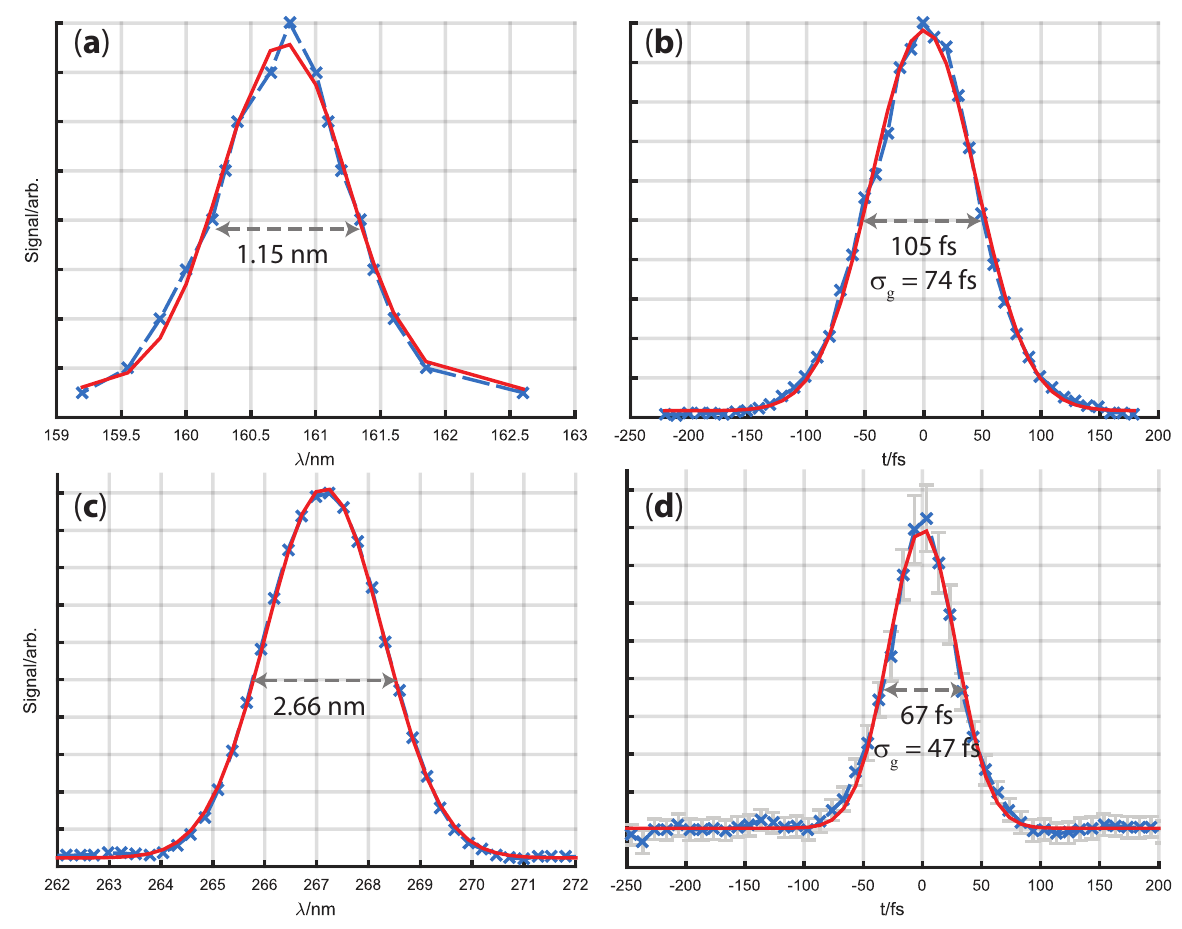}
\caption{{\label{fig:XCorr}(a) Spectrum of the 5$\omega$ pulse, measured with a VUV monochromator (McPherson model 234/302) and VUV photodiode (OSI Optoelectronics XUV 100). (b) Cross-correlation of the 5$\omega$ and 3$\omega$ pulses, measured via 1+1$'$ ionization of Xe in the VMI chamber. In both cases Gaussian curves (red) are fitted to the data (blue), and the full-width half-maxima are given; for the cross-correlation, this corresponds to pulse durations of $\sim$74~fs, under the assumption that both pulses are identical in duration. (c) Spectrum of the 3$\omega$ pulse, measured with an Ocean Optics HR400 spectrometer. (d) Autocorrelation of the maximally compressed 3$\omega$ pulse, measured outside of the experimental chamber with a 2-photon absorption autocorrelator \protect\cite{Homann_2011}.%
}}
\end{center}
\end{figure}

The laser, an amplified Coherent Legend Elite Duo, delivered 35~fs pulses at 800~nm with a pulse energy of 7.0~mJ at a repetition rate of 1~kHz. A 3.25~mJ component of the total laser output was used for the experiments detailed here, and Fig. \ref{fig:VUV_generation} provides a schematic illustration of part of this optical set-up. The beam was split into two arms, one of which (0.75~mJ) was frequency tripled using a pair of BBO crystals to generate the probe light (267~nm, 5~$\mu$J per pulse). The other arm was further split into two arms (the two input beams shown to the right of Fig. \ref{fig:VUV_generation}): the reflected component (1.5~mJ) was frequency tripled using the same scheme outlined above (``THG" in Fig. \ref{fig:VUV_generation} and shown in detail in the insert); and the transmitted component (1~mJ) remained at the fundamental frequency. The residual 400~nm and 800~nm light was separated from the 267~nm light by a series of mirrors with high reflectivity at 267~nm (HR 267nm). The pump light (160.8~nm, $\approx$0.5~$\mu$J) was generated by a non-collinear four-wave mixing scheme first demonstrated and described in detail by Noack and coworkers \cite{Beutler_2010,Ghotbi_2010}. The scheme makes use of a four-wave difference frequency mixing process in Ar, and was achieved by focusing the third harmonic and the fundamental into a gas cell containing Ar held at 32~Torr using curved high reflector mirrors: HR 267~nm ROC=1.5~m and HR 800~nm ROC=2~m, respectively.  Representative spectral and temporal data of the generated pulses are given in Fig. \ref{fig:XCorr}, and discussed further below.

Dichroic mirrors (Layertec GmbH) with high reflectivity at 160~nm (HR 160nm) and high transmission at both 267~nm and 800~nm were used for separating the VUV light and generating colors, and also for recombining the pump and probe beams in a collinear geometry. The collinear UV and VUV beams were both focused into the spectrometer using curved Al mirrors: ROC=1.4~m (CM) and ROC=2~m, for the pump and probe, respectively. The pump and probe beams were routed to the VMI spectrometer via a set of input baffles that minimized the transmission of scattered VUV/UV light to the laser-sample interaction region. Exit baffles were also incorporated to further reduce deleterious signals from scattered VUV/UV light. A variable time-delay between the pump and probe beams was introduced using a computer controlled delay stage (Newport XML210) and a set of retroflector mirrors. The temporal overlap between the pump and the probe pulses was determined $\emph{in-situ}$ by measuring non-resonant 1+1$'$ ionization in Xe, yielding a cross correlation of 105~fs (Fig. \ref{fig:XCorr}(b)). In Fig. \ref{fig:XCorr}(d) the $\emph{ex-situ}$ autocorrelation of the optimally compressed 3$\omega$ pulse is shown, as measured using an autocorrelator based on 2-photon absorption in bulk material \cite{Homann_2011}. This yielded a pulse duration of approximately 47~fs, close to the transform limit for the available bandwidth (40~fs). The VUV spectrum indicates a similar transform limit of 33~fs; however, due to material dispersion from the entrance/exit CaF$_{2}$/MgF$_{2}$ windows and Ar in the VUV generation cell, pulse durations on the order of 70-80~fs were expected for both the UV and VUV pulses in the interaction region of the VMI spectrometer, consistent with the measured $\emph{in-situ}$ cross-correlation (Fig. \ref{fig:XCorr}(b)).

\subsection{Velocity Map Imaging Spectrometer}

The VMI spectrometer consisted of a source and interaction chamber. The sample gases, typically 2\% of the molecule of interest seeded in He, were introduced into the spectrometer by means of a pulsed molecular beam. The beam was produced by a 1~kHz Even-Lavie pulse valve, which was heated to 35 $^{\circ}$C throughout the experiments. The molecular beam was expanded through a 250~$\mu$m conical nozzle into a source chamber typically held at a base pressure of 1$\times$10$^{-}$$^{6}$~Torr. The beam was then skimmed, to yield a beam with an estimated diameter of around 1~mm, before entering an interaction chamber along the spectrometer Time-of-Flight (ToF) axis, typically held at a base pressure of 1$\times$10$^{-}$$^{8}$~Torr, and intersected, at 90$^{\circ}$, by the co-propagating pump and probe laser pulses. Photoions produced from the pump-probe laser interaction were focused using ion optics onto a conventional MCP and phosphor-based detector setup and images were recorded using the PImMS camera.

The ion optic system incorporated a three-stage, open aperture repeller electrode system - similar to that described in Ref. \cite{Liu_2011} - that was used to minimize spurious signals from scattered VUV/UV light and background gas in the ionization chamber. The ion optic system additionally incorporated a set of eight Einzel lenses that allowed VMI or spatial imaging conditions to be achieved with a range of initial field gradients, facilitating tuning of the ion cloud compression/expansion along the ToF axis of the spectrometer. Following transit through the Einzel lens system, the ions were accelerated for detection using a high transmission grid electrode and detected using a pair of MCPs and a phosphor screen (P47), the emission from which was relay imaged to the PImMS sensor using an achromatic lens telescope. In the experiments reported here, a short flight tube was utilized with electrostatic lens voltages that maximized VMI resolution but encouraged charge cloud expansion along the ToF axis. This lead to a low - but adequate - mass-resolution of ~25~amu (FWHM) at 100~amu under the employed experimental conditions.

\subsection{PImMS Camera \& Data Acquisition}

The second generation PImMS2 sensor \cite{pimmsWeb} comprises a 324$\times$324 pixel array, with the ability to record up to four events per experimental cycle with 12.5~ns time resolution. The sensor can be read out at up to ~50~Hz. The latest versions of the PImMS2 sensor have been fitted with microlens arrays in order to direct incoming light towards the detection diodes in each pixel, thereby enhancing sensitivity. A more detailed overview of recent developments in fast sensors for time-of-flight imaging, including the PImMS camera, can be found in Ref. \cite{Vallance_2014}.

In the current experiments, photofragment images were recorded as a function of pump-probe delay in 20~fs steps, with 100 frames (laser shots) per step. Each sequence of delays - an experimental cycle - was kept short in order to minimize the effects of slow drifts over the cycle (primarily affecting the VUV power), but a number of cycles were run in order to build up good statistics. Typical datasets were obtained for a range of $\Delta t$ from -500 to +2000~fs, resulting in 125 time-steps, and requiring on the order of 1 hour to run several experimental cycles at a 50 Hz acquisition rate. After each experimental run (i.e. approximately every 1 - 2 hours) the VUV source was refreshed, by evacuating the source and refreshing the Ar gas, in order to maintain constant VUV output power. Further discussion of various effects in fs pump-probe configurations over long experimental timescales can be found in \cite{Hockett_2013}.

Data from the PImMS camera, in the form of $(x,y,ToF,f)$ quads, where $f$ is an index denoting the laser shot number within the multi-shot acquisition, was processed in Matlab using routines previously developed for data from coincidence imaging studies \cite{Hockett_2013}. Here $(x,y,ToF)$ provide the event position and time-of-flight information, and the laser-shot number, $f$, allows the data to be correlated with pump-probe delay $\Delta t$. From this raw data, various histograms and images can be obtained.

\section{Results}

\begin{figure}
\begin{center}
\includegraphics[width=1\columnwidth]{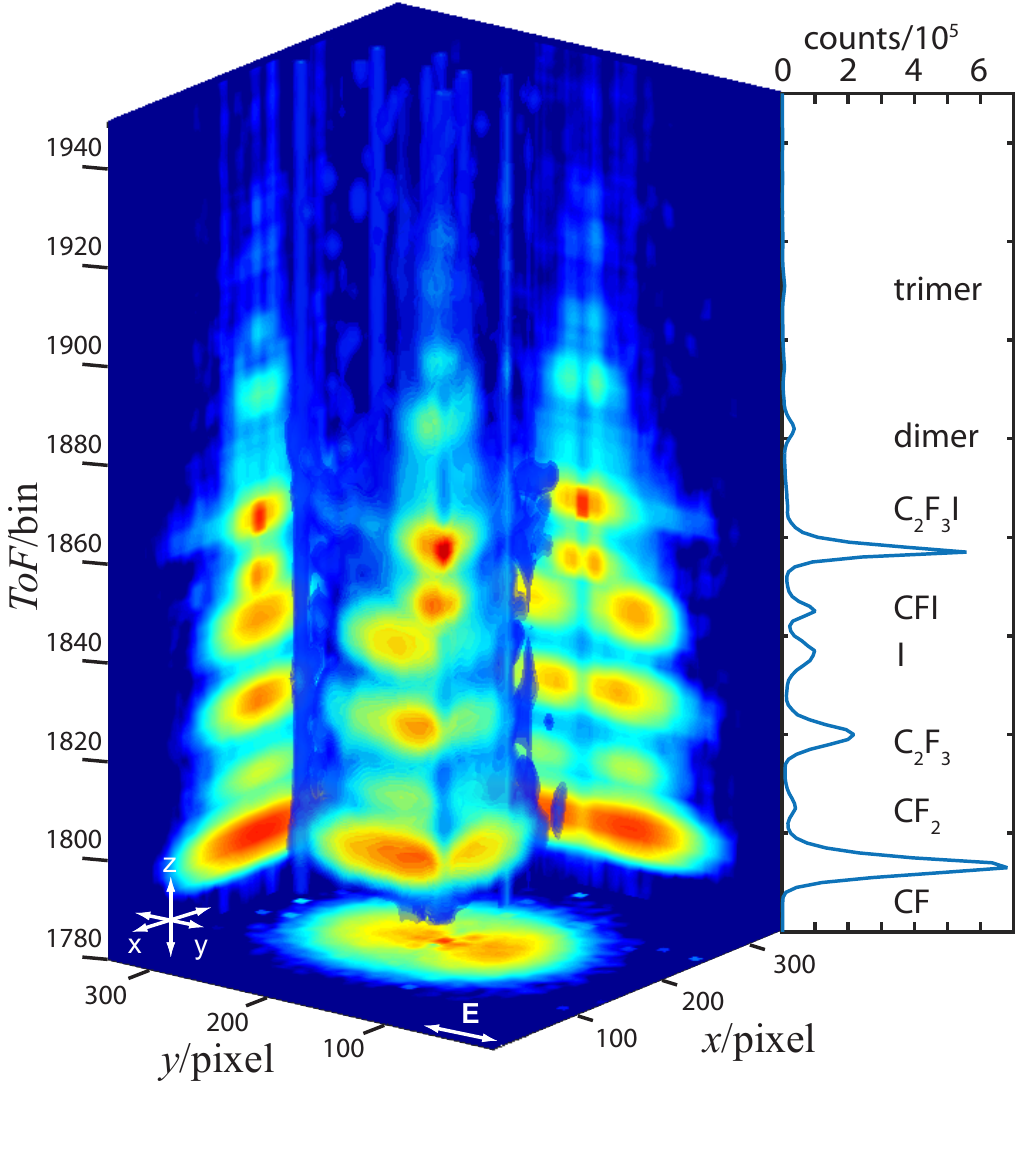}
\caption{{\label{fig:C2F3I-3Dtint}Time-integrated $(x,y,ToF)$ data from C$_2$F$_3$I, 8.4x10$^6$ events. The volumetric plot, shown for just one quarter of the dataspace, shows the full data as isosurfaces rebinned onto a 3D grid (25~ns per bin for ToF data), with a log10 color map. 2D image planes are also shown.  The $ToF$ spectrum, obtained by $(x,y)$ integration, is shown on the right, with peak assignments.%
}}
\end{center}
\end{figure}

VUV-UV pump-probe dynamics were studied for the polyatomic molecule C$_2$F$_3$I. The presentation here in focuses on the multi-mass ion imaging capabilities of the PImMS camera and the information content available in such multiplexed measurements. C$_2$F$_3$I has an ionization potential (IP) of 9.55~eV \cite{Schander_1977}; at $\lambda=$160.8~nm and $\lambda'=$267~nm, for 1+1$'$ absorption $E_{h\nu}=$12.35~eV (an excess energy of 2.8~eV), and for 1+2$'$ absorption $E_{h\nu}=$17~eV (excess energy of 7.45~eV). Therefore, for our experimental configuration - a short flight-tube VMI with an estimated mass resolution of $\approx$ 25~amu (FWHM) and VUV+UV laser pulses - C$_2$F$_3$I presents an ideal candidate for multi-mass ion imaging: multiple dissociation channels are expected, observed and readily resolvable in our configuration, and ultrafast molecular dynamics are expected prior to dissociation.

An overview of the multi-mass imaging data is provided in Fig. \ref{fig:C2F3I-3Dtint}. In this presentation, the raw data has been rebinned onto a 100x100x100 $(x,y,ToF)$ grid, integrated over pump-probe delay, and smoothed. In total 8.4x10$^6$ events are shown. The main part of the plot shows the counts per voxel, with a log10 color scale, and isosurfaces running from 20-90\% of the maximum voxel value (counts). 2D projections onto the ($x$,$y$), ($x$,$t$) and ($y$,$t$) planes are also shown, where the $(x,y)$ plane corresponds to traditional 2D imaging, and the other two planes provide $x$ or $y$ histograms as a function of particle time-of-flight. In this case, multiple features are observed, corresponding to different mass fragments with distinct times of flight and KE release. The right panel shows the integrated data - the time-of-flight spectrum - along with ion fragment peak assignments.

\begin{figure}
\begin{center}
\includegraphics[width=1\columnwidth]{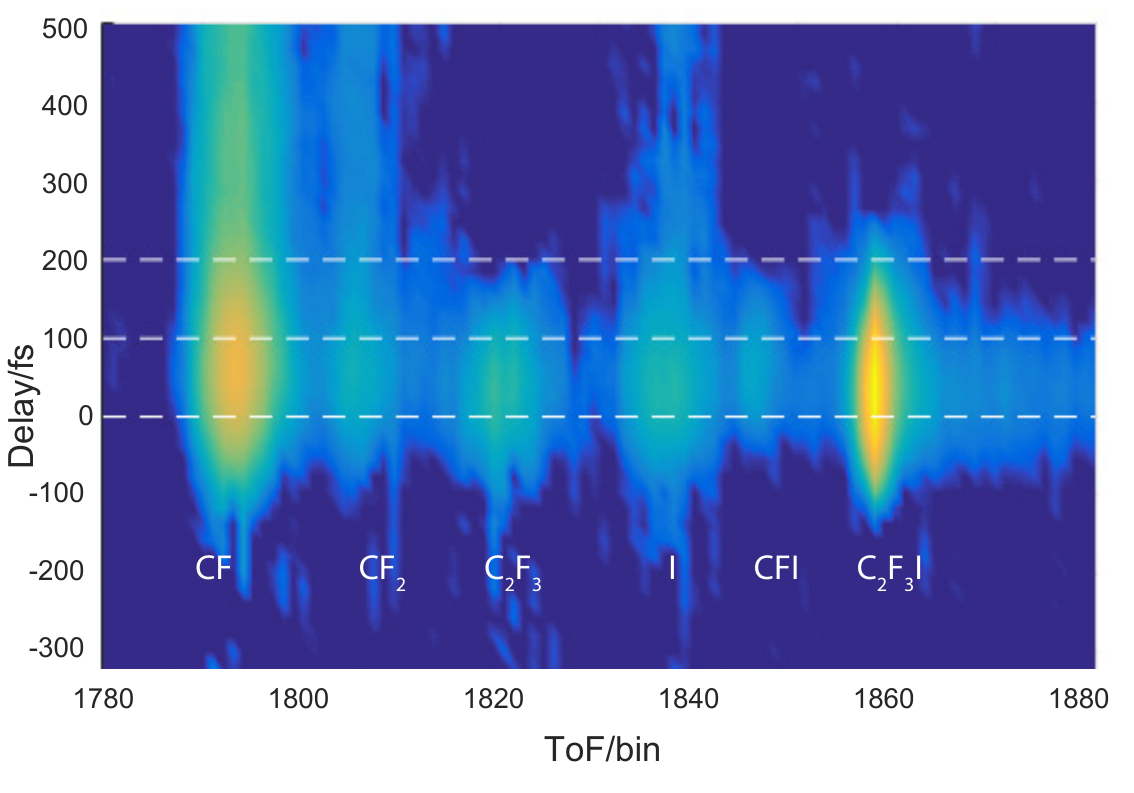}
\caption{{\label{fig:C2F3I-trTOF}Time-resolved data $(ToF,t)$ from C$_2$F$_3$I. In this plot, the data has been summed over $(x,y)$ and the background signals have been subtracted, to show the pump-probe time-dependence of each mass channel (log color map). Thresholding has been applied to the data to only display signal with values great than 0.05\% of the dataset maximum.  At $+ve$ delays the VUV pulse precedes the UV pulse, and vice versa for $-ve$ delays.%
}}
\end{center}
\end{figure}

\begin{figure}

\begin{center}
\includegraphics[width=1\columnwidth]{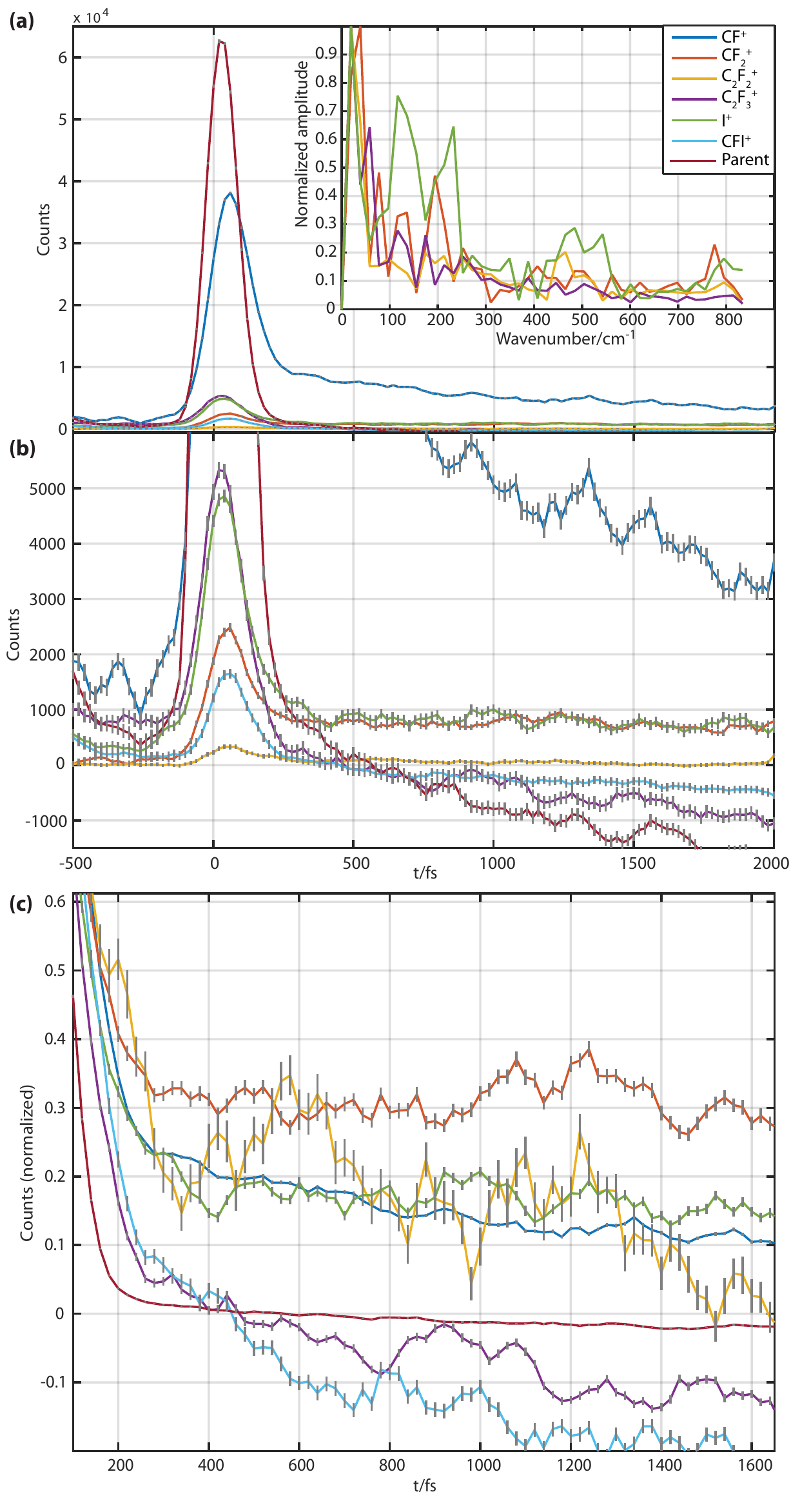}
\caption{{\label{fig:C2F3I-trTOF-lines}Time-resolved mass spectra from C$_2$F$_3$I. In this plot, the data from Fig. \ref{fig:C2F3I-trTOF} is further integrated over mass peaks (full width), to present a concise summary of the temporal dependence of each fragment channel. Background signals have been subtracted, and Poissonian uncertainties are shown. (a) Full dataset; the inset shows a Fourier transform of the time-series data for selected channels (omitted channels indicated no clear features from this dataset). (b) Detail of minor channels. Note that some over-subtraction of the background is clear with some channels exhibiting negative counts. (c) Normalized data over a reduced temporal range.%
}}
\end{center}
\end{figure}

An overview of the pump-probe time-dependence of the data is shown in Fig. \ref{fig:C2F3I-trTOF}. In this case, the data is binned onto a 2D grid $(ToF,t)$, to provide the time-resolved mass spectrum for all fragments, including some information on KE release along the ToF axis. By further integrating the data over $ToF$ windows defining each fragment, conventional time-resolved mass spectra are obtained, as shown in Fig. \ref{fig:C2F3I-trTOF-lines}. In this format, the counts per bin are shown directly, with the weakest fragment channel yielding just a few hundred events per delay at its peak.

From Figs. \ref{fig:C2F3I-3Dtint} - \ref{fig:C2F3I-trTOF-lines} the richness of a single dataset is clear. In Fig. \ref{fig:C2F3I-3Dtint} multiple fragments are observed, separated by $ToF$, each with distinct angular and KER distributions, which are clear in the full $(x,y,ToF)$ data. Overall, the delay-integrated data shows that CF$^+$ is the most abundant fragment observed, while C$_2$F$_3^+$, I$^+$ and the parent ion feature are all relatively close in total yield. In all cases the fragments are polarized parallel to the laser polarization, suggesting relatively fast fragmentation dynamics relative to the rotational timescale of parent. The CF$^+$ fragment shows a broad KER distribution, consistent with significant vibronic wavepacket dynamics in this pathway, leading to cleavage of the C=C bond; in contrast the C$_2$F$_3^+$ fragment shows a narrower KER distribution, consistent with a relatively fast and direct iodine loss channel. Weak dimer and trimer channels are also clearly observed, despite the low total counts for these channels (on the order of $10^2$ - $10^3$ events in total), indicating the utility and high dynamic range of the multi-mass ion imaging measurements.  

Fig. \ref{fig:C2F3I-trTOF} presents an overview of the time-dependence of the data, with binning on a $(ToF, \Delta t)$ grid. Additionally, background signal has been subtracted. In this case, no independent one-color (UV or VUV only) background signals were measured, so the background signal was approximated by averaging the early time data, over the range -500$<\Delta t<$-300~fs. This representation makes the overall temporal dynamics quite clear. A sharp rise and fall is observed for the parent ion signal, and temporally distinct behavior is observed for most of the fragment channels. In particular, the CF$^+$ fragment shows a delayed rise, and a long lifetime. There is the suggestion of oscillations in the tail, although this is not clear-cut in this visualization. 

Fig. \ref{fig:C2F3I-trTOF-lines} presents a more detailed view of the temporal dynamics of the fragment yields. In this case, the data is further integrated over the $ToF$ window for each fragment, to provide 1D data as a function of delay. In these plots, background signal has again been subtracted, and the data has additionally been smoothed with a 5-point moving average filter. 
Fig. \ref{fig:C2F3I-trTOF-lines}(a) shows the full dataset. In this plot, the temporal shift of the parent ion feature and the CF$^+$ feature is very clear, as is the distinct decay rates of the channels. Furthermore, the minor channels are also seen to show some clear differences in rise time and decay. Additional details can be extracted via a Fourier transform (FT) of the data, as shown in the inset. Although somewhat noisy, this analysis does indicate that characteristic frequency components are present in some of the mass channels, with the I$^+$ channel showing particularly clear features at $\sim$120 and $\sim$230~cm$^{-1}$, corresponding to oscillation periods of $\sim$280 and $\sim$140~fs respectively, and another broad feature centred at 500~cm$^{-1}$ (67~fs). The low frequency components are also observed in the CF$_2^+$ and C$_2$F$_3^+$ channel, suggesting correlated dynamics. 

Fig. \ref{fig:C2F3I-trTOF-lines}(b) shows the same data, on an increased scale, in order to show the details of the minor channels and longer time behavior. Here further details of the temporal dependence of the yields become apparent, and many of the channels appear to show oscillations. While these oscillations may appear noisy or chaotic, in most cases they are much larger than the Poissionian uncertainties. Furthermore, the correlations or anti-correlations observed in some channels suggest that the oscillations are likely to be genuine and physically reasonable. For instance, the parent and C$_2$F$_3^+$ channel appear anti-correlated, while the C$_2$F$_3^+$ and I$^+$ channels appear correlated. This observation is also consistent with the observation of shared characteristic frequencies in the FT data from these channels. However, it is clear that the approximate background subtraction is not effective at long delays, with the signal in some channels dropping significantly below zero counts. This is probably due to the time-dependence of the signal at negative delays, corresponding to 1$'$+1 or 2$'$+1 processes (i.e. UV pump, VUV probe), and this temporal dependence is additionally made obvious by oscillations in the CF channel at negative $\Delta t$. 

Finally, Fig. \ref{fig:C2F3I-trTOF-lines}(c) presents the same data, but normalized to the signal peak and on a magnified temporal axis. This further emphasizes the weakest fragment channels and again indicates that complex fragmentation dynamics are present. In particular, the minor C$_2$F$_2^+$ channel shows very clear features at 200, 600 and 900~fs, the latter of which appear to be anti-correlated with the CFI$^+$ channel; more rapid oscillations may also be present, but are difficult to discern. These time-scales are consistent with the small features in the FT for this channel, which indicate characteristic frequency components around 200 and 450~cm$^{-1}$, or 170 and 75~fs respectively. The CF$_2^+$ and I$^+$ channels also appear correlated, again a physically reasonable observation.


Overall, the time-resolved data indicates that the excited state, and its subsequent fragmentation, exhibits complex dynamics. There are likely to be multiple dissociation pathways for each observed fragment channel, and the dynamics on both the neutral and ionic surfaces may play a role. Different fragments show distinct decay lifetimes, and additionally small oscillations are observed in some channels, suggesting relatively localized or coherent vibrational wavepacket dynamics for some fragmentation pathways. The coarse temporal response and oscillations are closely correlated for the C$_2$F$_3^+$ and I$^+$, suggesting similar (or shared) wavepacket dynamics for these pathways, with charge localization on either moiety occuring close to dissociation. The CF$^+$ channel also shows very similar temporal dynamics, again suggesting similar (or shared) wavepacket dynamics for much of the dissociation pathway. The CF$_2^+$ and CFI$^+$ channels show slightly different behavior, with some counter-phased oscillations apparent to longer $\Delta t$, suggesting distinct wavepacket dynamics for these fragmentation channels. In contrast, the parent ion feature follows the pump-probe cross-correlation closely, with only slight temporal asymmetry and no significant long-lived contribution to the signal. However, while these temporal oscillations appear genuine and interesting, based on their scale relative to Poissonian uncertainties, the physically feasible correlations observed and the Fourier transformed data set, fully quantitative conclusions cannot yet be drawn from the current data. Further experiments to test for reproducibility, and obtain higher statistics, are required to verify and reinforce these observations. It is of note that analysis of single-cycle sub-sets of the data shown here revealed similar oscillations to the summed data presented in Figs. \ref{fig:C2F3I-3Dtint} - \ref{fig:C2F3I-trTOF-lines}, thus suggesting reproducibility.

\begin{figure*}
\begin{center}
\includegraphics[width=1\textwidth]{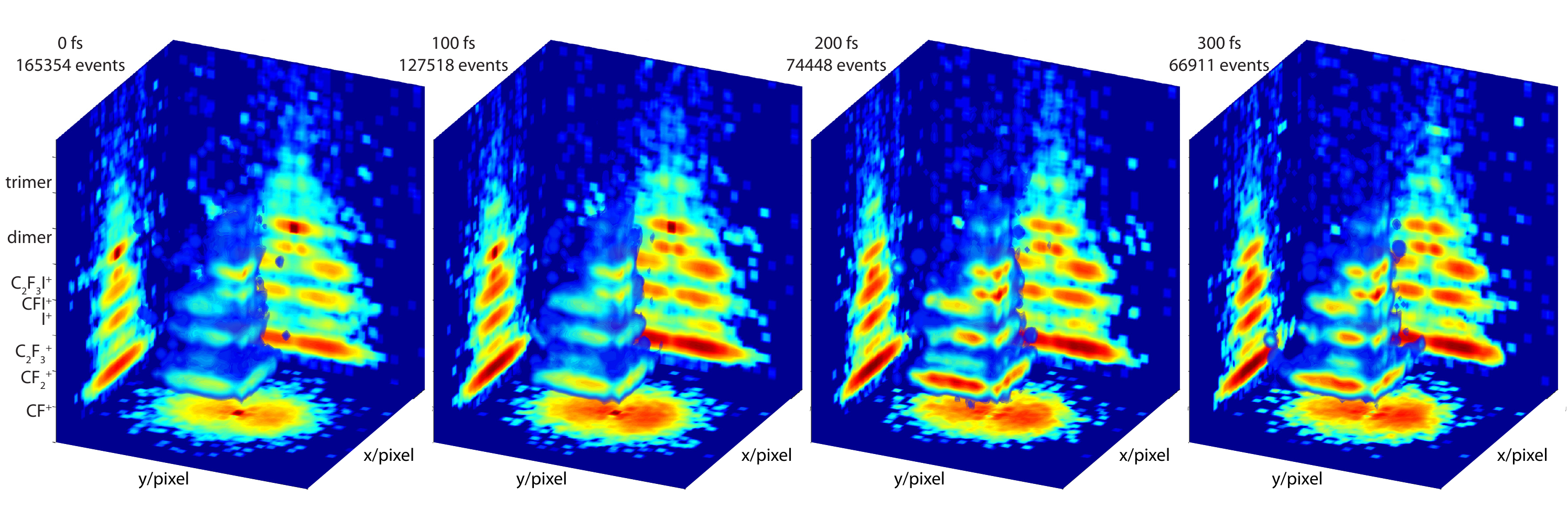}
\caption{{\label{fig:C2F3I-3Dt}Time-resolved $(x,y,ToF,\Delta t)$ data from C$_2$F$_3$I. The volumetric plots, shown for just one quarter of the dataspace, show the full data as isosurfaces rebinned onto a 3D grid (native bins), with a log10 color map. 2D image planes are also shown. Corresponding $(x,\Delta t)$ line-outs, with background subtraction, are shown in Fig. \ref{fig:tXdist}.%
}}
\end{center}
\end{figure*}

To further investigate the delay-dependent fragmentation dynamics phenomenologically, $(x,y,ToF,\Delta t)$ visualizations can be used. Fig. \ref{fig:C2F3I-3Dt} provides examples, in the same style as Fig. \ref{fig:C2F3I-3Dtint}. In this case, a few interesting features are observed. Firstly, underlying the direct parent ion signal - the intense feature in the center of the $(x,y)$ distribution, around $ToF$ bin \#1870 - is a broader feature, which is clearer at later delays where the parent ion spot is weak. This feature is most likely correlated with dimers which break-up rapidly in the acceleration region of the VMI spectrometer, yielding C$_2$F$_3$I$^+$ fragments with non-zero KE release. This is distinct from the main parent ion signal, which has zero KER, and hence appears as a narrow spot in the data, reflecting the parent velocity distributions within the molecular beam. The dimer signal at the parent mass appears invariant with $\Delta t$, and can be attributed primarily to 1-color only signal generated by the VUV laser. Secondly, there appears to be a gradual evolution in the KER distributions for some of the fragment channels; the CF$^+$ channel KER distribution sharpens slightly at later times, and the I$^+$ channel angular distribution becomes less polarized, with a broadening along the $ToF$ axis at later times. Thirdly, the CF$_2^+$ fragment shows hints of more complex time-dependence in its KER distribution; however, since the statistics are low this may purely be a result of rebinning and/or observational bias in these plots. 

\begin{figure}[h!]
\begin{center}
\includegraphics[width=1.0\columnwidth]{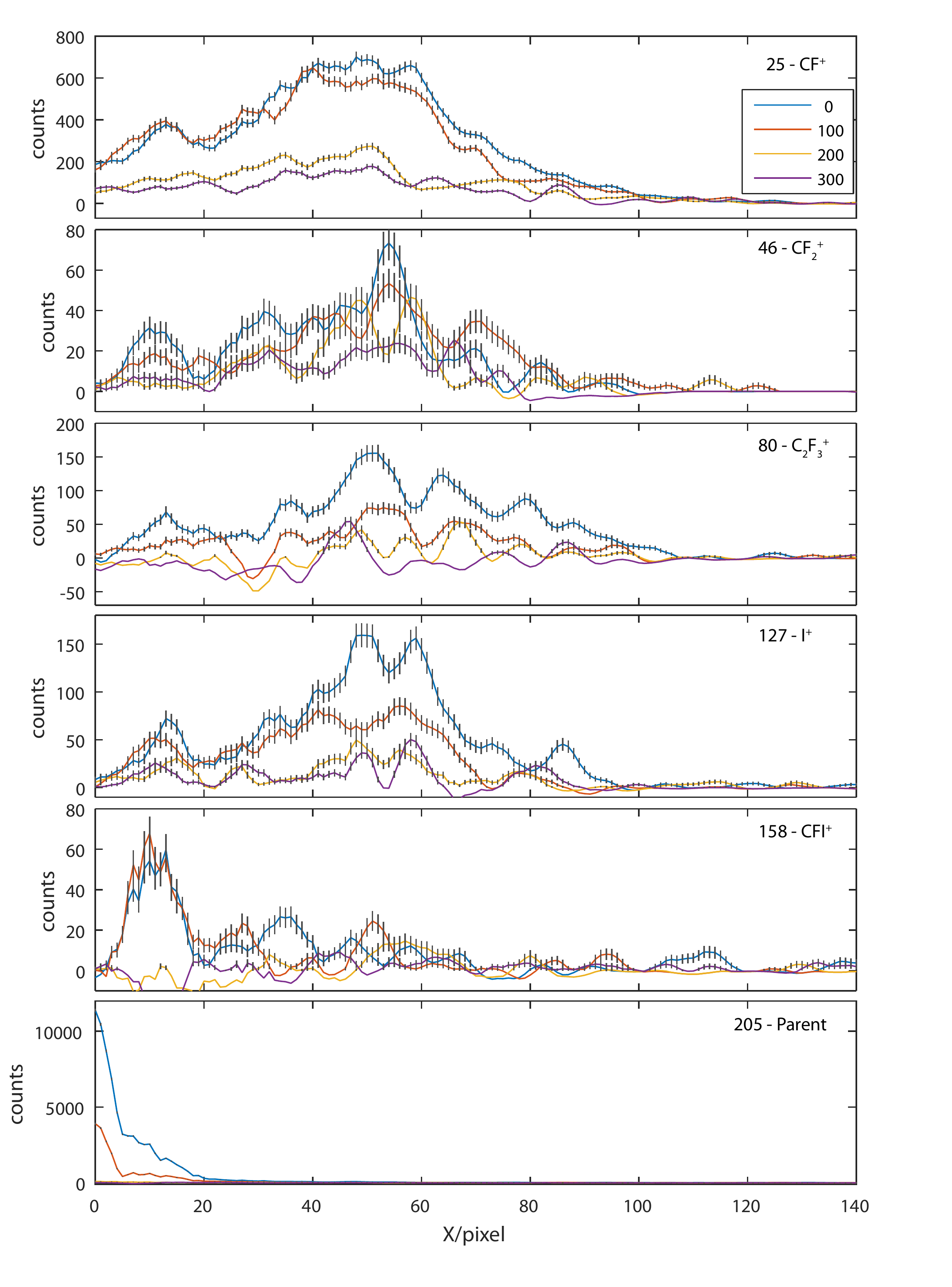}
\caption{{\label{fig:tXdist}Background-subtracted $(x,\Delta t)$ data, derived from the datasets shown in Fig. \ref{fig:C2F3I-3Dt}. Uncertainties are Poissonian, and the legend indicates $\Delta t$ for each trace in fs, $\pm10$. $x=0$ is chosen as the center of the distributions, in this case only half of the radial histogram is included in order to minimize the effect of detector inhomogeneities, which were severe over approximately one half of the MCP detector. Note that the apparent KER fine-structures present in some channels are likely unreliable at low counts ($<<10^4$), although the overall time-dependence is reliable. 
}}
\end{center}
\end{figure}

Specific aspects of the data can be investigated further by examining reduced dimensionality plots. Fig. \ref{fig:tXdist} provides an example, and shows the $(x,\Delta t)$ distributions corresponding to the full data shown in Fig. \ref{fig:C2F3I-3Dt}, with additional subtraction of background signals. Here, the data is histogrammed by $x$-pixel value at each delay $\Delta t$, providing an approximate mapping of the KER distribution along the laser polarization direction as a function of pump-probe delay. In this view, the observed behavior noted above can be seen more clearly, and additional features become apparent. A few specific examples are:
\begin{itemize}
\item The KER distribution for the CF$^+$ channel indeed sharpens slightly at later time-delays, with the signal edge moving from $x\approx$70 at $\Delta t=0$~fs to $x\approx$55 at $\Delta t$=200~fs. Additionally, the low energy feature at $x\approx$10 disappears at later times.
\item The I$^+$ channel shows a drop in the high-KER feature ($x\approx$90) as a function of time, and a low-KER feature ($x\approx$10) which is significantly reduced at later times. This observation is distinct from the observation of a less-polarized $(x,y,z)$ in the 3D plots at later times, but consistent with this behavior if there is little change in the total (angle-integrated) KER.
\item The CF$_2^+$ channel similarly shows a low-energy feature ($x\approx$10) which disappears at later times; additionally there are significant changes in the center of the distribution ($x$=50-60) apparent, with a narrow central feature at early times which appears to dissipate and, possibly, oscillate. However, given the low total counts present in this channel, this apparent fine-structure in the KER distribution (on the few pixel scale) and the temporal dependence of this fine-structure may be spurious.
\end{itemize}

As in the case of the fragment ion yields, the fragment kinetic energy release distributions present a detailed and complex picture of the temporal evolution of the excited-state wavepacket. Extensive further work is required in order to reach a quantitative understanding of the empirical observations made here; however, it is of note that changes in both the coarse and fine-structure, similar to those shown here on $\sim$100~fs timescales, were also found to occur on the $<$50~fs timescale, consistent with the typical timescales of vibrational wavepacket dynamics.

While the current data begins to present an interesting picture of the dynamics, elucidating precise details remain for further qualitative and quantitative analysis. This includes calibration of the data and determination of the $(V_x,V_y,V_z,\Delta t)$ KER distribution for each fragment, determination of the correlated angular distributions $I(\theta,\phi; |\mathbf{V}|,\Delta t)$, where $(\theta,\phi)$ define the ejection angle with respect to the laser polarization direction, and investigation of differences via subtraction of distributions at different $\Delta t$. Examples of this type of analysis can be found in Ref. \cite{Hockett_2013} for time-resolved data from butadiene, and Ref. \cite{Amini_2015} for UV photolysis of C$_2$H$_5$I in the energy domain. Additionally, further experiments will be required to probe the laser power dependence of the various fragment channels, thus allowing the separation of higher-order processes involving absorption of more than two photons. 

\section{Spectroscopy and Dynamics}

In terms of fundamental spectroscopy, C$_2$F$_3$I, along with other trifluoroetylene derivatives, was studied by Schander and Russell \cite{Schander_1977}. At 160.8~nm, the absorption spectrum consists of relatively sharp Rydberg features corresponding to $n\rightarrow6s$ and $\pi\rightarrow6p$ transitions, and an underlying broad quasi-continuum, assigned as the $n\rightarrow\pi*$ transition. Some vibrational structure was observed in the $p$ Rydberg features, and has been assigned to the C=C stretch of the ion (1530~cm$^{-1}$); the $s$ Rydbergs are higher-lying in energy and converge to the second ionization potential (11.2~eV). Based on this absorption spectrum, relatively complex wavepacket dynamics are expected for excitation with broadband VUV radiation. The dispersed spectrum of the $\pi*$ state indicates a short lifetime, hence strong coupling to lower-lying states, and long C=C vibrational progressions are expected, while the low symmetry of the species suggests that strong vibronic couplings and IVR will be present, consistent with the observed short lifetime. 

Based on C$_{2}$F$_{3}$I dissociation limits determined in previous VMI experiments \cite{NRC_VMI_200nm_experiments}, available dissociation limit/thermochemistry data \cite{Franck_1990,Caballero_1999,Bacskay_2015}, and the vibrational constants reported in the literature \cite{Geboes_2015,Wurfel_1991,McNaughton_1992,Sendt_2000,Bacskay_2015,Carrington_1970,McGurk_1973,Tiemann_1973}, the energetic fragmentation thresholds of C$_{2}$F$_{3}$I can be estimated. Of note here are the thresholds for C$_{2}$F$_{3}$ + I($^{2}$P$_{3/2}$) (3.20$\pm$0.05~eV), CF$_{2}$+CFI (4.6$\pm$0.1~eV), and CF$_{2}$ + CF + I($^{2}$P$_{3/2}$) (6.6$\pm$0.1~eV) production; all of which are accessible following single-photon excitation at 7.71~eV and potential signatures of which are observed in the data shown in Figs. \ref{fig:C2F3I-3Dtint}-\ref{fig:tXdist}. Building upon these neutral fragmentation thresholds using the known ionization potentials of the fragment species \cite{Lias1988,NIST_2006,Bieri_1980}, the dissociative ionization thresholds can also be estimated. Of note here are the C$_{2}$F$_{3}^{+}$ + I($^{2}$P$_{3/2}$) (13.4$\pm$0.2~eV), C$_{2}$F$_{3}$ + I$^{+}$($^{3}$P$_{2}$) (13.68$\pm$0.05~eV), CF$_{2}^{+}$ + CFI (16.1$\pm$0.2~eV), CF$_{2}$ + CFI$^{+}$ (unknown CFI ionization potential and, hence, dissociative ionization threshold), and CF$_{2}$ + CF$^{+}$ + I($^{2}$P$_{3/2}$) (15.74$\pm$0.15~eV) thresholds. Considering the photon energies utilized in these experiments, all of these dissociative ionization limits can be reached via 1+2$'$ excitation processes. Considering the neutral and ionic fragmentation limits and the observed ion mass peaks , we note that the transient ion signals presented in Figs. \ref{fig:C2F3I-trTOF}-\ref{fig:tXdist} could be produced via neutral fragmentation and multi-probe-photon, fragment ionization or alternative dissociative ionization pathways. Given that the transient ion signals are generally observed to decay as a function of pump-probe delay, it is suggested that the latter mechanism is predominantly responsible for the fragment ion data reported here. However, extended pump-probe delay scans and probe power signal dependence studies would be required to affirm this inference; both of which are deferred at this point.

To the best of our knowledge, no time-resolved spectroscopy and photodissociation studies of C$_2$F$_3$I in the VUV have previously been published. However, direct (single-photon) photoionization and fragmentation studies of the similar species C$_2$H$_2$F$_3$I (trifluoroethyl iodide, IP=10.0~eV \cite{Watanabe_1962}) in the VUV have recently been published, at photon energies from 10-22~eV \cite{Lago_2013}. In that work, photoelectron-photoion coincidence measurements were made, providing a detailed mapping of the fragmentation products as a function of wavelength. The I$^+$ product was observed at energies as low as 12.13~eV, but significant other fragmentation products were not observed until higher energies, $>$13~eV (see Fig. 4 and table 2 of Ref. \cite{Lago_2013} for further details). 

In this vein, it is also of note that the richness and complexity of excited state dynamics in the somewhat similar cases of ethylene \cite{Tao_2011, Allison_2012} and butadiene \cite{Levine_2009} pumped by UV/VUV photons have been considered in detail with the aid of ab initio dynamics calculations. The studies on ethylene are of particular relevance to the current work. In those studies, time-resolved pump-probe spectroscopy was performed with $\sim$161~nm pump pulses, and various high-harmonic probe wavelengths. Time-resolved ion yields were recorded, and ab initio multiple spawning (AIMS) dynamics calculations were performed. In that work multiple fragmentation channels were observed, along with complex time-dependent dynamics conceptually analogous to those observed herein for C$_2$F$_3$I. In particular, CH$_3^+$ and CH$_2^+$ fragments were observed with complex temporal behavior, while the H$_2^+$ and H$^+$ channels appeared to grow smoothly as a function of time, and to dominate at long delays. The authors concluded that ``The experimental data and the AIMS simulations indicate the presence of fast, non-statistical, elimination channels for H$_2$ molecules and H atoms in the photolysis of C$_2$H$_4$." In this case, the excitation at 160~nm is $\pi\rightarrow\pi*$, while in C$_2$F$_3$I the ``perfluoro effect" results in a lowering of the $\pi*$ state \cite{Schander_1977}, and the absorption spectrum indicates a dominant  $n\rightarrow\pi*$ transition as discussed above; the descent in symmetry from D$_{2h}$ to C$_s$ implies more strongly coupled vibronic states. Nonetheless, and rather broadly speaking, somewhat similar classes of dynamics may still be expected, but with iodine elimination and C=C fission playing a key role, due to the additional element of the ``perfluoro effect" weakening the C=C bond \cite{Lin_1998}.

While less directly relevant to the case herein, the butadiene AIMS calculations revealed the complex charge-localization dynamics that may appear transiently with wavepacket motion in polyatomic molecules. In that case, positive charge was localized on the backbone, or terminal CH$_3$ group, resulting in two relaxation pathways with similar vibrational dynamics, but  distinct charge localization dynamics. In the case of dissociative ionization, these pathways would correlate with different fragmentation patterns, although for the case of slow dissociation timescales further dynamics on the ionic surfaces may obviate this strong correlation. Additionally, dynamics including C=C bond alternation and terminal group twisting and wagging were investigated; such dynamics were seen to occur on characteristic timescales of 20-50~fs and, in dissociative ionization measurements, would result in  significant changes in fragment speed and angular distributions as a function of time-delay. While these dynamics are specific to butadiene, again the general characteristics or classes of behavior might be expected to be somewhat ubiquitous for similar molecules, since complex wavepacket dynamics underlie both the excited state dynamics and fragmentation dynamics (either on the neutral or ionic surfaces). The coincidence imaging studies discussed previously, which are conceptually similar to the multi-mass VMI studies herein, were able to discern changes to the photoelectron angular distributions and ion fragment KER distributions on these timescales \cite{Hockett_2013}.

\section{Discussion \& Conclusions}
The data presented herein indicate the types of datasets that will soon be routinely obtainable when combining the PImMS multi-mass imaging camera with ultrafast pump-probe spectroscopy. In the work reported here, a single 3D multi-mass imaging dataset with on the order of 10$^6$ events could be obtained in $\sim$1 hour of experimental time. This can be contrasted with alternative single or few particle imaging methodologies based on delay-line detectors, currently the only other means of performing equivalent 3D multi-mass imaging. Such detectors are typically limited to a few events per laser shot, resulting in event acquisition rates close to the laser repetition rate. For kHz laser systems, this can be restrictive, and lead to very long experimental times. For example, the butadiene coincidence studies of Ref. \cite{Hockett_2013}, required approximately 48 - 72 hours of data acquisition for a similar number of events, although the count rate was further restricted to a few hundred Hz in that case to maintain coincidence conditions. While it is now feasible to run such long experiments, which necessitate extremely stable experimental and environmental conditions (typically requiring a combination of passive and active methods to obtain), they are not routine and also do not scale well with the number of datasets desired: quantitative pump-probe power-dependence studies are not
feasible, for example.

Conceptually similar, but experimentally quite different, 3D ion data can also be obtained with gated VMI type configurations. In this case, different ion masses are obtained sequentially, via detector gating over a $ToF$ region of interest. Typically 2D images are obtained, from which the full 3D distribution can be easily reconstructed if cylindrical symmetry is maintained. A careful comparison of this type of conventional ion imaging with PImMS has previously been made \cite{Amini_2015}. For full 3D imaging without the requirement of cylindrical symmetry, one can also employ tomographic techniques \cite{Wollenhaupt_2009,Smeenk_2009,Hockett_2010}. This works well, but requires yet more sequential data acquisition - specifically the recording of multiple 2D images as a function of pump-probe laser polarization with respect to the detector (projection) plane. Taking the C$_2$F$_3$I example herein, for an arbitrary pump-probe polarization geometry one would require 6 mass windows, and 8 projections per window (assuming a maximum of $L$=4 in the angular distributions, and Nyquist sampling in the number of projection angles required) at each pump-probe delay. In this case, one would therefore require 48 2D images per $\Delta t$. Thus, for multi-mass imaging of complex fragmentation patterns as a function of pump-probe delay, this methodology quickly becomes impractical. For further discussion on these topics, the reader is referred to Chichinin et. al. for general 3D VMI discussion \cite{Chichinin_2009}, Ref. \cite{Rolles_2007} for discussion of multi-hit delay-line anodes for VMI, the works of Li and co-workers for the use of fast frame cameras in conjunction with a high-speed digitizers \cite{Lee_2014a,Lin_2015,Winney_2016}, and Vallance $\emph{et. al}$. for general discussion of fast sensors for VMI \cite{Vallance_2014}.

In terms of probing ultrafast molecular dynamics in polyatomic molecules, the capabilities and benefits of multi-mass ion imaging are clear. Furthermore, the use of VUV sources provides a route to 1+1$'$ pump-probe experiments in a larger range of molecules, and a means to increase the observation window for dynamics probed via one-photon ionization \cite{Fuji_2011,Kobayashi_2015}. While the analysis of the data and determination of the specific details of the dynamics in a given case remains a formidable experimental and theoretical challenge, more complete experimental datasets are a necessary starting point. It is of note that with the development of computational methodologies such as AIMS, and ongoing increases in computer power, $\emph{ab initio}$ studies are gradually becoming routine \cite{Levine_2008,Tao_2009,Mori_2012}. In future work we plan to extend these capabilities with both higher-order processes (6th and 7th harmonics of the fundamental 800~nm driving field) and tunable VUV, by obtaining more detailed multi-mass ion data (higher statistics, power dependence, polarization geometry dependence), and via PImMS-VMI modalities allowing for multi-mass ion images to be obtained in coincidence or covariance with electron images, extending previous work such as refs. \cite{Davies_1999,Gessner_2006,Lin_2015,Lehmann_2012}.

\section{Acknowledgements}

We are grateful to Andrey Boguslavskiy and Martin Larsen for assistance with the experimental infrastructure, and to Denis Guay and Doug Moffat for technical support. AS acknowledges the NSERC Discovery Grant program for financial support. JWLL, MB, MB and CV gratefully acknowledge financial support from the EPSRC via programme grant EP/L005913/1, the EU (FP7 ITN 'ICONIC', Project No. 238671), and the STFC (PNPAS award and mini-IPS grant No. ST/J002895/1).

\bibliographystyle{apsrev4-1}
\bibliography{converted_to_latex}

\begin{thebibliography}{60}
\bibitem [{NRC()}]{NRC_VMI_200nm_experiments}%
  \BibitemOpen
  \href@noop {} {\enquote {\bibinfo {title} {The
  {C$_{2}$F$_{3}$I}$\rightarrow${C$_{2}$F$_{3}$} + {I($^{2}$P$_{3/2}$)}
  dissociation limit was extracted from {VMI} data recorded in previous single
  photon dissociation, {I($^{2}$P$_{3/2}$)} fragment-selective (2+1)
  resonantly-enhanced multiphoton ionization {VMI} experiments at the nrc
  laboratory. by carefully calibrating the {I($^{2}$P$_{3/2}$)} kinetic energy
  release using data recorded in similar experiments performed with a
  {CH$_{3}$I} sample and the {CH$_{3}$I} dissociation limit reported in {Ref.}
  62, a dissociation limit of 3.20$\pm$0.05~{eV} was determined for
  {C$_{2}$F$_{3}$I}.}}\ }\BibitemShut {NoStop}%
\end{thebibliography}

\end{document}